\documentclass[aps,pre,twocolumn,groupedaddress,longbibliography,nofootinbib,showkeys,showpacs]{revtex4-1}

\usepackage{amsmath}
\usepackage{amsfonts}
\usepackage{graphicx}
\graphicspath{ {./} }
\usepackage{mathptmx}      
\usepackage{bm}
\usepackage{enumerate}
\usepackage{subfigure}
\usepackage{color}
\usepackage{array}
\usepackage{tabularx}
\usepackage{lipsum}
\usepackage{tikz}

\newcommand{\ket}[1]{| #1 \rangle} 

\definecolor{DarkGreen}{RGB}{0,100,0}
\begin{document}


\title{General error mitigation for quantum circuits}

\author{Manpreet Singh Jattana}
\affiliation{Institute for Advanced Simulation, J{\"u}lich Supercomputing Centre, Forschungszentrum J{\"u}lich, D-52425 J{\"u}lich, Germany}
\affiliation{RWTH Aachen University, D-52062 Aachen, Germany}

\author{Fengping Jin}
\affiliation{Institute for Advanced Simulation, J{\"u}lich Supercomputing Centre,\\ Forschungszentrum J{\"u}lich, D-52425 J{\"u}lich, Germany}
\author{Hans De Raedt}
\affiliation{Institute for Advanced Simulation, J{\"u}lich Supercomputing Centre, Forschungszentrum J{\"u}lich, D-52425 J{\"u}lich, Germany}
\affiliation{Zernike Institute for Advanced Materials,\\
University of Groningen,
NL-9747 AG Groningen, The Netherlands}
\author{Kristel Michielsen}
\affiliation{Institute for Advanced Simulation, J{\"u}lich Supercomputing Centre,\\ Forschungszentrum J{\"u}lich, D-52425 J{\"u}lich, Germany}
\affiliation{RWTH Aachen University, D-52062 Aachen, Germany}
\date{\today}

\begin{abstract}
A general method to mitigate the effect of errors in quantum circuits is outlined. The method is developed in sight of characteristics that an ideal method should possess and to ameliorate an existing method which only mitigates state preparation and measurement errors. The method is tested on different IBM Q quantum devices, using randomly generated circuits with up to four qubits. A large majority of results show significant error mitigation. 
\end{abstract}

\maketitle

\section{Introduction}\label{INT}
The road of developing and operating devices that would enable quantum computation has been and continues to be full of obstacles. While some of the development obstacles had been pointed early on, i.e. implementation of reversibility \cite{doi:10.1098/rsta.1995.0106} and loss of coherence \cite{PhysRevA.51.992}, some are found as we progress. Despite these, devices had been developed and small problems have been implemented \cite{Kandala2017}. The implementations bring with them operational obstacles. One operational obstacle is the presence of both known (e.g.~noise, decoherence) and unknown causes that render the computation erroneous. To tackle this obstacle to some extent, error correction had been proposed \cite{shorec, PhysRevLett.77.3260, PhysRevA.57.127}. Error correction is difficult to implement on current devices due to its hardware requirements. Another attempt at removing the erroneous computation obstacle is error mitigation. In this article we propose and test a new method for it.

\subsection{Correction versus mitigation}
  Error correction has been described as a procedure of protecting quantum computation against noise or errors, by encoding redundant information to the process \cite{Nielsen}. This redundancy necessitates additional hardware in order to be implementable. If we wish to avoid using additional hardware, we need a different method altogether. 

  If we define `cooperative' as the ability of a quantum device to systematically reproduce its errors, then, \textit{error mitigation} is defined as a method to attenuate errors when applied to a cooperative error prone device. Since such a process does not require additional hardware, but only additional resources from the available hardware, the main problem that error correction faces is circumvented.
  
  Henceforth, we focus on error mitigation.

\subsection{Requirements}\label{sec:req}
  We require a method to have the following characteristics to be called an ideal error mitigation method:
  \begin{enumerate}
  \item Result recovery: the method should be able to mitigate errors to a satisfactory accuracy.
  \item Depth independence: the method should not depend on circuit depth.
  \item Error model: the method should take into account all types of errors a device may be prone to, and not rely on prior information about errors the device is prone to.
  \item Practically realisable: the method should make use of resources that are practically similar to the resources used by the circuit which is to be (error) mitigated.
  \item No additional hardware: the method should not require additional quantum hardware for being implementable. 
  \item Gate-set independence: the method should take into account and be applicable to all kinds of quantum gates. 
  \item No output knowledge: the method should not make use of any specific knowledge about the output of a given circuit.
  \end{enumerate}

\subsection{Mitigation methods}
  Various error mitigation methods have been introduced recently. These include methods for error extrapolation and probabilistic error mitigation \cite{PhysRevLett.119.180509, Li2017, PhysRevX.8.031027, Strikis}, methods that utilise symmetries in circuits and use it for certain errors elimination \cite{McArdle2019,PhysRevA.98.062339, PhysRevA.100.010302}, and ideas like quantum subspace expansion \cite{McClean2017} or decoherence-free subspaces \cite{Premakumar}, quantum process tomography \cite{Bialczak2010, Howard2006, Chow2009, Neeley2008}, gate set tomography \cite{Merkel2013, Greenbaum2015, Blume-Kohout2017} and quasi-probability decomposition \cite{Songeaaw5686}. There are methods that focus on read-out error mitigation based on detector tomography \cite{Maciejewski}. Methods have been tested on trapped-ion devices \cite{Zhang} and superconducting devices \cite{Kandala2019, Songeaaw5686}. 
  
We focus on a standard error mitigation method for state preparation and measurement (SPAM) errors mitigation (SPAMEM), which is available in IBM's Qiskit library \cite{Qiskit}. This method fulfils requirements 2, 5, and 7 and uses a matrix based approach to mitigate errors. We confine ourselves to comparing the herein proposed method with Qiskit's standard method. The herein proposed method satisfies the requirements 2, 3, 5, 6, and 7. It satisfies 1 in most of the cases we tested. Requirement 4 is discussed later (see Sect.~\ref{sec:rrc}). This new method builds upon Qiskit's method and is not limited to SPAM but mitigates all kinds of errors. 

\section{Qiskit error mitigation}\label{sec:qemod}
Two types of errors produced in any error prone device can be assumed to be coming from state preparation and measurement. If a large part of an erroneous device output is due to SPAM, which is not known a priori, SPAMEM is a useful method. Under the assumption that SPAM errors for a given circuit $C_g$ will also occur for other circuit(s) $C_c$, we may mitigate them by measuring the effects produced by SPAM errors if the outputs of $C_c$ were known. We call, therefore, $C_c$ a calibration circuit.

  Assume that we have an error prone device which produces some relative frequencies ($v_1, v_2, ..., v_{2^N}$), which differ from the ideal (exact) data ($e_1, e_2, ..., e_{2^N}$). Later on, when applying error mitigation, we will require a notation for mitigated data ($x_1, x_2, ..., x_{2^N}$) and simulator data ($s_1, s_2, ..., s_{2^N}$) as well, so let us construct the column vectors
  \begin{equation}\label{eqn:1}
  V = \begin{pmatrix}
v_1 \\
v_2 \\
  \vdots \\
v_{2^N} \\
  \end{pmatrix},\quad
  E = \begin{pmatrix}
e_1 \\
e_2 \\
  \vdots \\
e_{2^N} \\
  \end{pmatrix},\quad
  X = \begin{pmatrix}
x_1 \\
x_2 \\
  \vdots \\
x_{2^N} \\
  \end{pmatrix}  ,\quad
    S = \begin{pmatrix}
s_1 \\
s_2 \\
  \vdots \\
s_{2^N} \\
  \end{pmatrix}
 .
  \end{equation}

  All these vectors are normalised, i.e. $\sum v_i=\sum e_i=\sum x_i=\sum s_i=1$, where $i=1,...,2^N$. Now we postulate the existence of a $2^N \times 2^N$ matrix $M$ such that
  \begin{equation}
  ME =V.\label{eqn:m}
  \end{equation}
  Equation~(\ref{eqn:m}) serves as a good starting point to understand the basic idea of the method. Note that if the device is not error prone, then $M$ is the identity matrix. For an error prone device, $M$ has non-zero off-diagonal elements. The purpose of a mitigation method, in the context of this article, is to propose a procedure to be performed on the device, take the outputs to be put in $M$ and use it for error elimination. In the next section, we outline the procedure.

\subsection{Calibration and mitigation}\label{sec:calandmit}
  To start, we fix the number of qubits and denote it by $N$. We will require $2^N$ circuits to be run for the mitigation. The procedure is as follows:
  \begin{enumerate}
  \item Prepare the qubits in all possible $2^N$ states and measure each state. Each such circuit is a calibration circuit.
  \item Enter the data from each calibration circuit into columns of $M$ (and rename it $M_Q$), where the $j^{th}$ column, starting from left, takes data from the circuit whose state is given by the binary representation of $j$, for all $j=1,...,2^N$. We refer to $M_Q$ as the \textit{Qiskit calibration matrix}.
  \end{enumerate}
  Now that we have $M_Q$, let us use this to mitigate (errors in) the relative frequencies $V$ for a circuit $C_g$. Instead of Eq.~(\ref{eqn:m}), we now have $M_QX=V$, where $X$ is the mitigated data and may not be always exact (i.e. equal to $E$). When we proceed to solve for $X$, we face a problem. Since we did not take into account the fact that we are dealing with relative frequencies which are constrained to be in the interval [0,1], by simply using the inverse of $M_Q$ to solve for $X$, we may have values in $X$ outside this interval \cite{Maciejewski, Geller2020}. To avoid this problem, we use least squares. Thus, we change our problem to finding the minimum of the function
  \begin{equation}\label{eqn:fx}
   f(x)= \sum^{2^N}_{i=1} (v_i - (M_Q \cdot X)_i)^2,
  \end{equation}
  given the constraints  $0\leq x_i \leq 1$  and $\sum x_i=1$ for all $i=1,...,2^N$. For our experiments we initialize $X$ randomly and use the \textit{minimize} package of \textit{scipy} \cite{scipy} and the \textit{Sequential Least SQuares Programming} (SLSQP) \cite{Kraft} method.
 
It is important to note that although Eq.~(\ref{eqn:m}) cannot be used in some cases, using Eq.~(\ref{eqn:fx}) is always possible. This makes the mitigation method a useful heuristic \cite{Geller2020}.

\subsection{Discussion}
  We now have the error mitigated data $X$ and if we want to test the method, we can compare it to the data $S$ produced by a simulator (see Eq.~(\ref{eqn:1})). Let us introduce the root mean square errors
  \begin{equation}\begin{split}\label{eqn:def1}
  & \Delta X = \sqrt{\sum_{k=1}^{2^N} \big(x_k - s_k\big)^2}
  \text{ and } \Delta V = \sqrt{\sum_{k=1}^{2^N} \big(v_k - s_k\big)^2}.
  \end{split}
  \end{equation}
 If we further define $\Delta_Q = \Delta V - \Delta X$, we can have the following possibilities:
\begin{equation}\Delta_Q 
\begin{cases}\label{eqn:poss}
    >0 ,& \text{for positive mitigation.} \\
    <0, &  \text{for negative mitigation. }\\
    =0, & \text{for no mitigation.} 
\end{cases}
\end{equation}

According to requirement 1 from Sect.~\ref{sec:req}, $\Delta_Q$ should be positive for all experiments. Furthermore, the level of `satisfaction' addressed in requirement 1 can be quantified in terms of $\Delta_Q$. For a positive mitigation to be perfect, $\Delta X = 0$ when $\Delta V \neq 0$. 

Applying this method on the real device for small depth ($D\approx 2$) circuits works well (data not shown). For larger circuit depths, where significant errors can come from gate operations, Qiskit error mitigation does not improve the results considerably, as shown later (see Sect.~\ref{sec:qem}). In the next section, we therefore propose a general error mitigation method.

\section{General Error Mitigation (GEM)}
If we add some gates to the preparation and measurement circuit, and the gate operations are error prone, SPAM errors are no longer the only sources of errors. In cases where SPAM errors are not dominant, the Qiskit mitigation approach is unsuitable. In general, we require to mitigate effects of errors that arise not only due to SPAM, but potentially also due to other sources, including but not limited to (for example) erroneous gate operations. If we further postulate that we do not know \textit{apriori} what sources will contribute to or dominate the errors, as might be the case for any practical quantum computation, we need to include all potential sources of errors. To this end, we need a general method to mitigate errors in a quantum circuit. 

For a circuit of any given depth, gate-set, and number of qubits, errors can come from multiple sources. While some sources may be identified and modelled, some others remain unidentified. The identification and the accurate modelling of the error sources present a major challenge. For the general mitigation method, we do not attempt to take this challenge and take a different route, as explained below.

Recollecting our assumption from the beginning of Sect.~\ref{sec:qemod} and applying it to a general case we conclude: under the assumption that all errors in a given circuit $C_g$ will also occur for other circuit(s) $C_c$, we may mitigate them by measuring the effects these errors produced if the outputs of $C_c$ were known. Using this approach we avoid modelling individual errors in a circuit altogether and work directly with the device.

For consistency, we borrow Eqs.~(\ref{eqn:1}) and (\ref{eqn:fx}) and proceed directly by proposing the procedure to be performed on the device to produce $M$.

\subsection{Calibration and mitigation}

Assume that we wish to mitigate errors in a given circuit $C_g$ of depth $D$ and with $N$ qubits. The procedure is as follows:
\begin{enumerate}
 \item Prepare calibration circuits in all possible $2^N$ states, twice.
 \item Consider all the gates of the circuit $C_g$ up to depth $D/2$ (if $D$ is even) or $(D-1)/2$ (if $D$ is odd), and add them to the calibration circuits.
 \item Add \textit{inverse gates} of the gates added in step 2, in a reverse order.
 \item Measure the calibration circuits and record the data in calibration matrix $M_1$, similarly as done in Sect.~\ref{sec:calandmit} step 2.
 \item Repeat steps 2, 3, and 4 for the remaining half of the gates on the remaining calibration circuits, and name the new matrix $M_2$.
 \item Calculate the matrix $M_G=(M_1+M_2)/2$.
\end{enumerate}
Now, we may proceed by using Eq.~(\ref{eqn:fx}) to find $X$ by simply replacing $M_Q$ by $M_G$.

\subsection{Example}
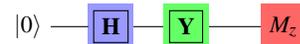
\begin{figure}[h]
\begin{tikzpicture}\centering
\node at (-0.8,0.3) {\textbf{$\ket{0}$}};
\fill[blue!40] (0,0) rectangle (0.6, 0.6);
\node[draw] at (0.3,0.3) {\textbf{H}};
\draw (0.6,0.3) -- (1,0.3);
\draw (-0.5,0.3) -- (0,0.3);
\draw (2.3,0.3) -- (1.6,0.3);
\fill[green!60] (1,0) rectangle (1.6, 0.6);
\node[draw] at (1.3,0.3) {\textbf{Y}};
\fill[red!60] (2.3,0) rectangle (2.9, 0.6);
\node at (2.6,0.3) {\textbf{$M_z$}};
\end{tikzpicture} \caption{Circuit showing a \textbf{H} (Hadamard) and \textbf{Y} gate followed by a measurement in the Z basis.\label{fig:example}}
\end{figure}

To illustrate the GEM procedure, let us consider a simple circuit with one qubit, as shown in Fig.~\ref{fig:example}.

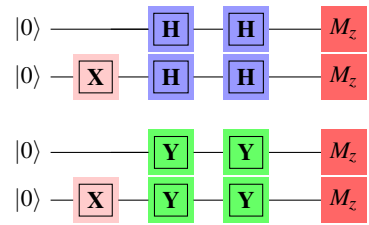
\begin{figure}[h]
\begin{tikzpicture}
\node at (-1.6,0.3) {\textbf{$\ket{0}$}};

\fill[blue!40] (0,0) rectangle (0.6, 0.6);
\node[draw] at (0.3,0.3) {\textbf{H}};
\draw (0.6,0.3) -- (1,0.3);
\draw (0,0.3) -- (-1.3,0.3);
\draw (-0.5,0.3) -- (0,0.3);
\draw (2.3,0.3) -- (1.6,0.3);
\fill[blue!40] (1,0) rectangle (1.6, 0.6);
\node[draw] at (1.3,0.3) {\textbf{H}};
\fill[red!60] (2.3,0) rectangle (2.9, 0.6);
\node at (2.6,0.3) {\textbf{$M_z$}};
\end{tikzpicture}

\begin{tikzpicture}
\node at (-1.6,0.3) {\textbf{$\ket{0}$}};
\fill[red!20] (-1,0) rectangle (-0.4, 0.6);
\node[draw] at (-0.7,0.3) {\textbf{X}};
\fill[blue!40] (0,0) rectangle (0.6, 0.6);
\node[draw] at (0.3,0.3) {\textbf{H}};
\draw (0.6,0.3) -- (1,0.3);
\draw (-1,0.3) -- (-1.3,0.3);
\draw (-0.4,0.3) -- (0,0.3);
\draw (2.3,0.3) -- (1.6,0.3);
\fill[blue!40] (1,0) rectangle (1.6, 0.6);
\node[draw] at (1.3,0.3) {\textbf{H}};
\fill[red!60] (2.3,0) rectangle (2.9, 0.6);
\node at (2.6,0.3) {\textbf{$M_z$}};
\end{tikzpicture}

\begin{tikzpicture}
\fill[red!00] (-1,0) rectangle (-0.4, 0.3);
\fill[red!00] (2,0) rectangle (-0.4, 0.3);
\end{tikzpicture}\begin{tikzpicture}
\fill[red!00] (-1,0) rectangle (-0.4, 0.3);
\fill[red!00] (2,0) rectangle (-0.4, 0.3);
\end{tikzpicture}

\begin{tikzpicture}
\node at (-1.6,0.3) {\textbf{$\ket{0}$}};
\fill[green!60] (0,0) rectangle (0.6, 0.6);
\node[draw] at (0.3,0.3) {\textbf{Y}};
\draw (0.6,0.3) -- (1,0.3);
\draw (0,0.3) -- (-1.3,0.3);
\draw (-0.5,0.3) -- (0,0.3);
\draw (2.3,0.3) -- (1.6,0.3);
\fill[green!60] (1,0) rectangle (1.6, 0.6);
\node[draw] at (1.3,0.3) {\textbf{Y}};
\fill[red!60] (2.3,0) rectangle (2.9, 0.6);
\node at (2.6,0.3) {\textbf{$M_z$}};
\end{tikzpicture}

\begin{tikzpicture}
\node at (-1.6,0.3) {\textbf{$\ket{0}$}};
\fill[red!20] (-1,0) rectangle (-0.4, 0.6);
\node[draw] at (-0.7,0.3) {\textbf{X}};
\fill[green!60] (0,0) rectangle (0.6, 0.6);
\node[draw] at (0.3,0.3) {\textbf{Y}};
\draw (-1,0.3) -- (-1.3,0.3);
\draw (0.6,0.3) -- (1,0.3);
\draw (-0.4,0.3) -- (0,0.3);
\draw (2.3,0.3) -- (1.6,0.3);
\fill[green!60] (1,0) rectangle (1.6, 0.6);
\node[draw] at (1.3,0.3) {\textbf{Y}};
\fill[red!60] (2.3,0) rectangle (2.9, 0.6);
\node at (2.6,0.3) {\textbf{$M_z$}};
\end{tikzpicture}
 \caption{Four different calibration circuits for the circuit shown in Fig.~\ref{fig:example}.\label{fig:cal} We prepare the circuits twice in the states $\ket{0}$ and $\ket{1}$ and construct identity circuits by splitting the circuit in half and adding each half together with its inverse (see text).}
\end{figure}

The corresponding calibration circuits are depicted in Fig.~\ref{fig:cal}.
The outputs obtained from measuring the calibration circuits are then put into matrices $M_1$ and $M_2$.
\begin{equation}
 M_1 = \begin{pmatrix}
  v_1 & v_3\\
 v_2 & v_4  \\
  \end{pmatrix},
  \quad
   M_2 = \begin{pmatrix}
  v_5 & v_7\\
  v_6 & v_8  \\
  \end{pmatrix}\label{eqn:m1m2}.
\end{equation}
The left column of $M_1$ in Eq.~(\ref{eqn:m1m2}) is (the output obtained) from the first circuit in Fig.~\ref{fig:cal}, the right column from the second, the left column of $M_2$ from the third, and the right column from the last circuit, respectively. The final matrix $M_G$ is then simply the average of $M_1$ and $M_2$.
\section{Experimental results}
\subsection{Randomised testing}
The GEM method is applied to randomly generated circuits for a large number of cases. Although the method is independent of the gate-set used, for practical reasons we apply only a subset of all possible gates. The used gate-set reads: \{\textbf{Id}, \textbf{U1}, \textbf{X}, \textbf{Y}, \textbf{Z}, \textbf{H}, \textbf{S}, \textbf{S{$^\dagger$}}, \textbf{T}, \textbf{T{$^\dagger$}}, \textbf{CNOT}\}. These gates are transpiled into device compatible gates as shown in Table~\ref{table:2}.
\begin{table}
\caption{The left columns shows the gates that were added to the generated circuits and the right column shows their corresponding basis gates implemented on the quantum device.}\label{table:2}
\begin{tabular}{|c|c|}
\hline
Applied Gate & Basis Gate  \\ \hline
\textbf{ID}&\textbf{ID}\\ \hline
\textbf{U1}($\theta$) &\textbf{U1} ($\theta$)  \\ \hline
 \textbf{X}&\textbf{U3}($\pi,0,\pi$)  \\ \hline
 \textbf{Y}&\textbf{U3}($\pi,\pi/2,\pi/2$)  \\ \hline
 \textbf{Z}& \textbf{U1}($\pi$) \\ \hline
 \textbf{H}& \textbf{U2}($0, \pi$) \\ \hline
 \textbf{S}& \textbf{U1}($\pi/2$) \\ \hline
 \textbf{S{$^\dagger$}}& \textbf{U1}($-\pi/2$) \\ \hline
  \textbf{T}& \textbf{U1}($\pi/4$) \\ \hline
   \textbf{T{$^\dagger$}}& \textbf{U1}($-\pi/4$) \\ \hline
\textbf{CNOT(c$\to$t)}&\textbf{CNOT(c$\to$t)}\\ \hline
   \end{tabular}
\end{table}
Note that on all the different devices tested by us, where all gates are decomposed into some elementary gates called ``basis gates'', we used all the basis gates. The gate-set we used is a universal gate-set. Therefore, the tested quantum circuits are non-trivial. We created
a one-dimensional array of all gates in our gate-set and used \textit{NumPy choice} \cite{10.5555/2886196} to randomly choose gates to be added to the circuits. To assess the proposed method, we test it on the IBM Q devices for different numbers of qubits and circuit depths. The following points were kept in mind:
\begin{figure}

\begin{tikzpicture}
\node[inner sep=0pt] (russell) at (0,0)
    {\includegraphics[trim = {0.55cmcm 0.95cm 0.5cm 0.2cm},scale=0.55]{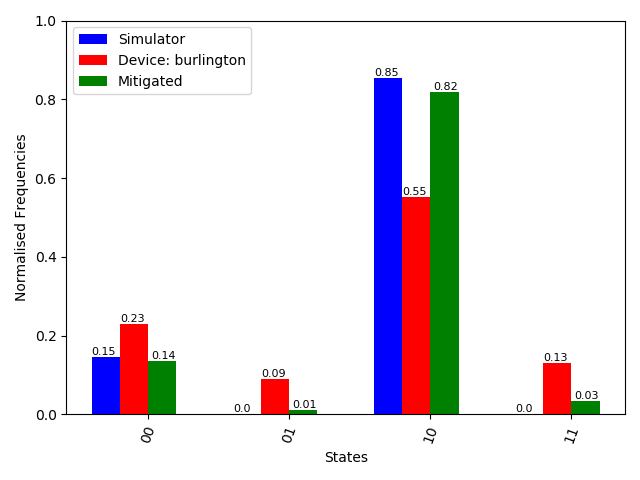}};
\fill [color=white] (-2.1,2.425) rectangle (-1,2.17) ;
   \node at (-1.55,2.29) {\scriptsize Burlington};
\end{tikzpicture}
\caption{A sample experiment using the general error mitigation (GEM) method. Shown here are the simulator (left; blue), the device (middle; red), and mitigated (right; green) results for one of the repetitions of experiment number 79 in Fig.~\ref{fig:3}.}\label{fig:bar1} 
\end{figure}
\begin{figure}
    \includegraphics[trim = {0.55cm 0.9cm 0.5cm 0.5cm},scale=0.555]{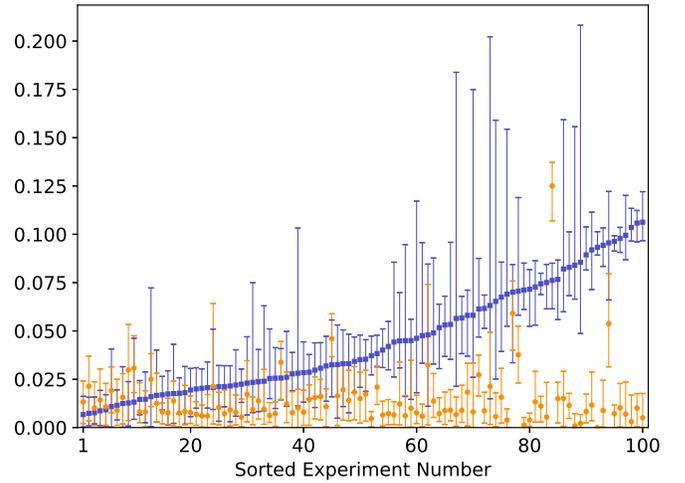}
    \caption{The average of $\Delta V$ (blue square points) and the corresponding average of $\Delta X$ (orange circular points) for $100$ random circuits, each repeated $10$ times, with $N$=1 (qubit) and depth $D \in [20,29]$ with mean depth $\bar{D}=24.73$. The top and bottom of the error bars for $\Delta V$ and $\Delta X$ represent the maximum and minimum obtained in 10 repetitions, respectively. The device used was IBM Q Armonk \cite{dev:armonk}. Data is presented in ascending order of average $\Delta V$ for easy readability.}\label{fig:1}
\end{figure}
\begin{figure}
    \includegraphics[trim = {0.55cm 0.9cm 0.5cm 0.5cm},scale=0.555]{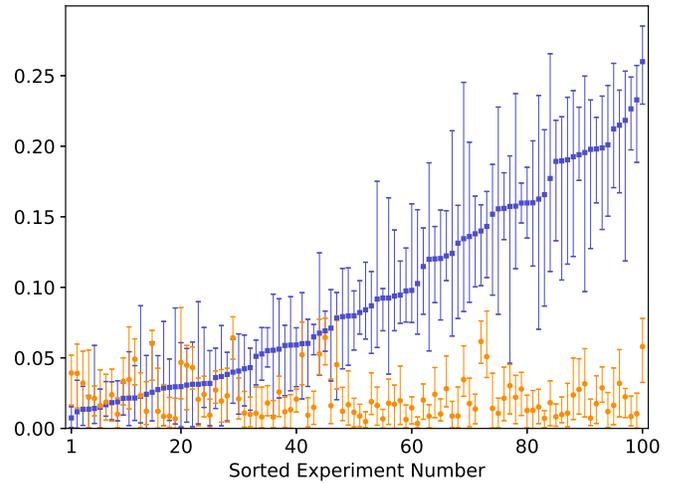}
    \caption{Same as Fig.~\ref{fig:1} except for $D \in [89,112]$ with $\bar{D}=99.26$. }\label{fig:1b}
\end{figure}
\begin{figure}
    \includegraphics[trim = {0.55cm 0.9cm 0.5cm 0.5cm},scale=0.555]{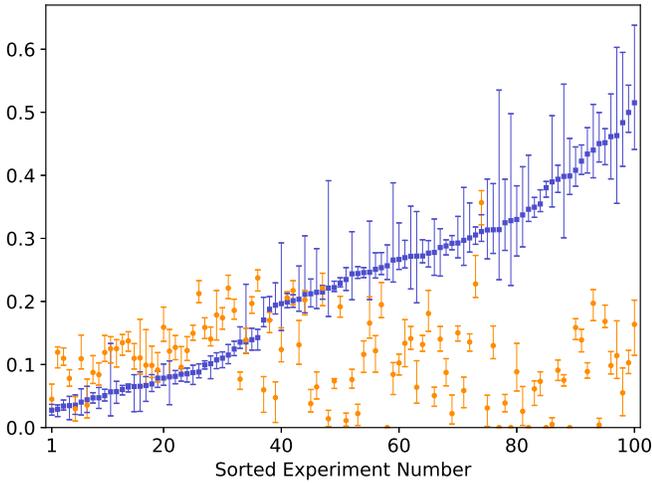}
    \caption{Same as Fig.~\ref{fig:1} except for $N = 2$ and $D \in [16,20]$ with $\bar{D}=19.50$. The device used was IBM Q Burlington \cite{dev:burlington}.}\label{fig:3}
\end{figure}
\begin{figure}
    \includegraphics[trim = {0.55cm 0.9cm 0.5cm 0.5cm},scale=0.555]{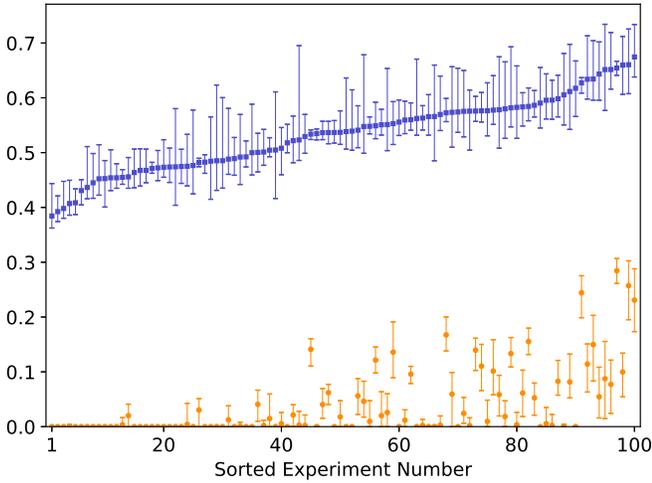}
    \caption{Same as Fig.~\ref{fig:1} except for $N = 2$ and $D \in [74,80]$ with $\bar{D}=79.38$. The \textbf{H} gate was not used in the gate-set. The device used was IBM Q Burlington \cite{dev:burlington}. }\label{fig:2}
\end{figure}

\begin{figure}
    \includegraphics[trim = {0.55cm 0.9cm 0.5cm 0.5cm},scale=0.555]{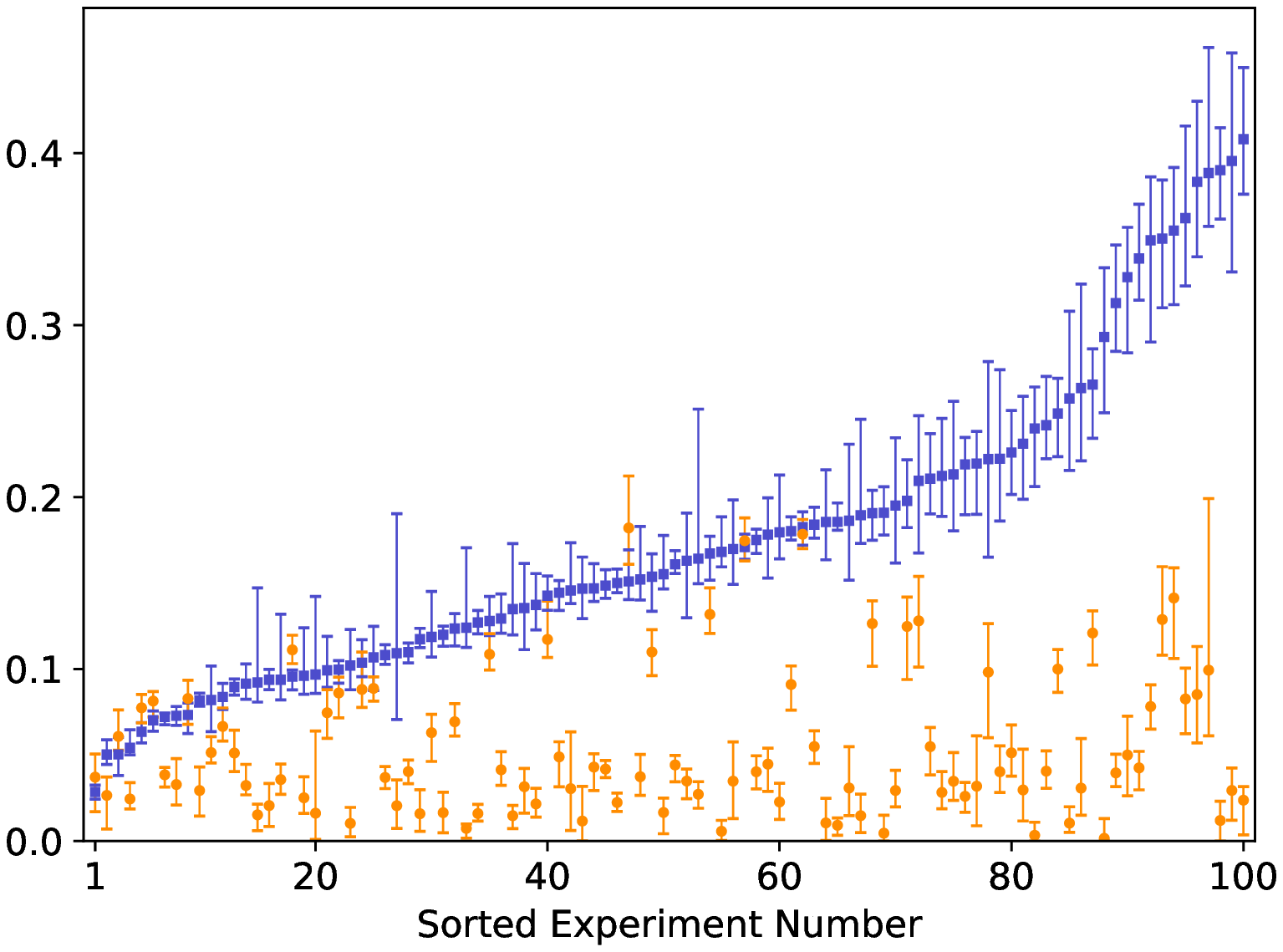}
    \caption{Same as Fig.~\ref{fig:1} except for $N = 3$ and $D \in [6,10]$ with $\bar{D}=8.32$. The device used was IBM Q Ourense \cite{dev:ourense}.}\label{fig:4}
\end{figure}

\begin{figure}[ht]
    \includegraphics[trim = {0.55cm 0.9cm 0.5cm 0.5cm},scale=0.555]{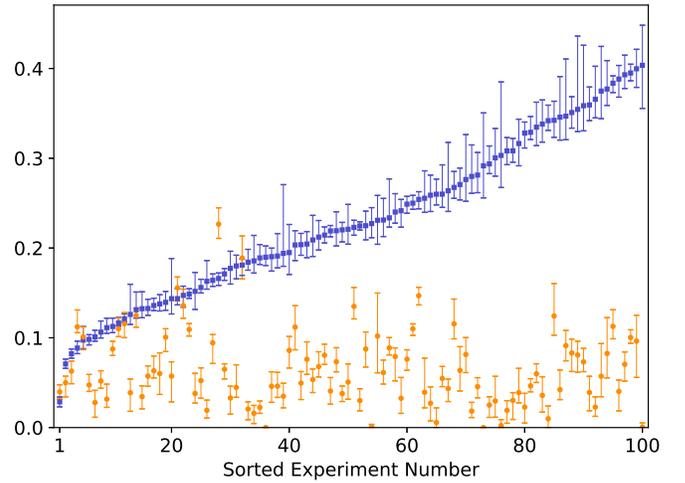}
    \caption{Same as Fig.~\ref{fig:1} except for $N = 4$ and $D \in [6,10]$ with $\bar{D}=7.06$. The device used was IBM Q Ourense \cite{dev:ourense}.}\label{fig:6}
\end{figure}
\begin{itemize}
 \item Create random circuits respecting the physical connectivity of the qubits on the device. Repeat each circuit and its mitigation, to observe any statistical deviations. Repetitions were limited to ten times for all experiments.  
 \item The depths $D$ for each $N$ were chosen after finding the suitable depths for error mitigation, keeping in mind the actual device performance based on $\Delta V$. 
 \item Although different devices perform differently, no preference was given to any particular combination of $N$ and $D$ to any device. Total shots for all experiments were kept at the maximum supported by the devices, which was $8192$ per experiment. The simulator was run with $819200$ shots. Care should be taken in comparing results for different $N$ and $D$ because they may not have been obtained from the same device. 
 \item We set the \textit{circuit optimization level} in Qiskit to zero, and do not include the measurement (gate) in the circuit depth. 
\end{itemize}

\subsection{Results and discussion}
Figure~\ref{fig:bar1} shows one example of how the output looks like for one experiment on the simulator, on the device, and after error mitigation. For this example, GEM closely recovers the theoretical (exact) results. In this work, in total, we performed six hundred different experiments each repeated ten times. Results are shown in Figs.~\ref{fig:1}--\ref{fig:6}, where we plot the average $\Delta V$ and $\Delta X$ (see Eq.~(\ref{eqn:def1})) for the ten repetitions of each experiment along with the maximum and minimum datum obtained. By using error mitigation we are hoping to reduce $\Delta V$, where the new value (called the mitigated value) is given by $\Delta X$. Positive values of $\Delta V$ (blue squares) represent an erroneous output, and lower values of $\Delta X$ (orange circles) represent a more successful mitigation, where $\Delta X=0$ is a perfect mitigation giving the exact result. For ease of readability, we plot the data in ascending order of the average $\Delta V$.

In Figs.~\ref{fig:1} and \ref{fig:1b}, we performed GEM on one qubit and different mean circuit depths $\bar{D}\approx 25$ and $\bar{D}\approx 99$, respectively. The effect of increased depth is visible in Fig.~\ref{fig:1b} where values of $\Delta V$ are much larger than those in Fig.~\ref{fig:1}. This was expected since a larger number of error prone gates contribute more errors in total. Note that in both cases the error bars are large, suggesting that the output of an experiment repeated ten times fluctuates significantly. 

In Figs.~\ref{fig:3} and \ref{fig:2}, we performed GEM on two qubits and different mean circuit depths $\bar{D}\approx 20$ and $\bar{D}\approx 78$, respectively. In the $\bar{D}\approx 20$ case, negative mitigation occurs significantly more often and this is addressed in the forthcoming subsection (see Sect.~\ref{sec1}). In the $\bar{D}\approx 78$ case, we deliberately remove the \textbf{H} gate from the gate set, so that we have outputs that yield only one state, where the effect of errors is most pronounced. We observe positive mitigation for all experiments.

In Fig.~\ref{fig:4}, we performed GEM on circuits with three qubits and mean circuit depth $\bar{D}\approx 8$. We see that to observe similar $\Delta V$ for higher qubit numbers, the depth has to be decreased accordingly and sometimes significantly. This is also visible in Fig.~\ref{fig:6}, where we performed GEM on circuits with four qubits.

We borrow the definitions from Eq.~(\ref{eqn:def1}) and define $\Delta_G = \Delta V - \Delta X$ for the general case. Also for GEM, $\Delta_G$ covers all the possibilities given in Eq.~(\ref{eqn:poss}). Let us now analyse the data in terms of $\Delta_G$. We choose all $\Delta_G<|0.03\times \text{max}(\text{average}(\Delta V))|$, or $3\%$ of largest average deviation, to be $\Delta_G\approx 0$. Table~\ref{table} shows the number of experiments, corresponding to each $N$ and $\bar{D}$, for which we observed positive, negative, or no mitigation. For most experiments, positive mitigation is seen.
\begin{table}[]
\caption{The distribution of $\Delta_G$ observed in the experiments for different qubits and mean depths (see caption of corresponding figure). $\Delta_G>0$ is positive mitigation, $\Delta_G<0$ is negative mitigation, and $\Delta_G\approx 0$ is no mitigation.\label{table}}
\begin{tabular}{cccc}
\\ \cline{1-4} 
\multicolumn{1}{|c|}{Figure} & \multicolumn{1}{c|}{$\Delta_G>0$} & \multicolumn{1}{c|}{$\Delta_G<0$} & \multicolumn{1}{c|}{$\Delta_G\approx0$} \\ \cline{1-4} 
\multicolumn{1}{|c|}{\ref{fig:1}} & \multicolumn{1}{c|}{85} & \multicolumn{1}{c|}{11} & \multicolumn{1}{c|}{4} \\ \cline{1-4} 
\multicolumn{1}{|c|}{\ref{fig:1b}} & \multicolumn{1}{c|}{81} & \multicolumn{1}{c|}{12} & \multicolumn{1}{c|}{7} \\ \cline{1-4} 
\multicolumn{1}{|c|}{\ref{fig:3}} & \multicolumn{1}{c|}{60} & \multicolumn{1}{c|}{32} & \multicolumn{1}{c|}{8} \\ \cline{1-4} 
\multicolumn{1}{|c|}{\ref{fig:2}} & \multicolumn{1}{c|}{100} & \multicolumn{1}{c|}{0} & \multicolumn{1}{c|}{0} \\ \cline{1-4} 
\multicolumn{1}{|c|}{\ref{fig:4}} & \multicolumn{1}{c|}{91} & \multicolumn{1}{c|}{3} & \multicolumn{1}{c|}{6} \\ \cline{1-4} 
\multicolumn{1}{|c|}{\ref{fig:6}} & \multicolumn{1}{c|}{90} & \multicolumn{1}{c|}{2} & \multicolumn{1}{c|}{8} \\ \cline{1-4} 
\end{tabular}
\end{table}
\subsubsection{GEM versus QEM}\label{sec:qem}
\begin{figure}
    \includegraphics[trim = {0.55cm 0.9cm 0.5cm 0.5cm},scale=0.555]{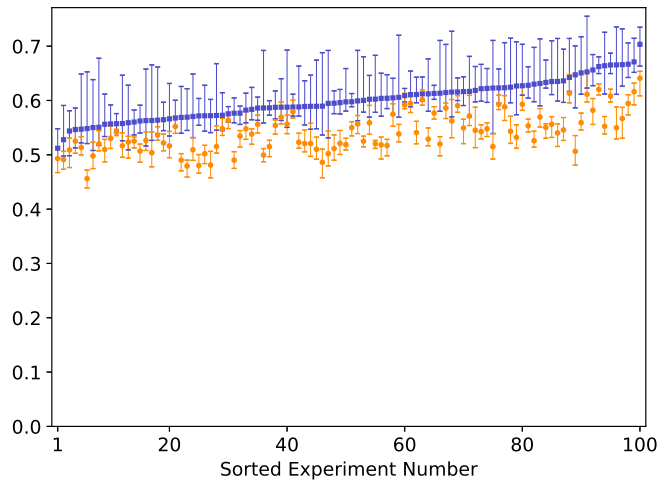}
    \caption{Same as Fig.~\ref{fig:2} except that Qiskit error mitigation was used for $D \in [72,80]$ with $\bar{D}= 77.50$. The device used was IBM Q Burlington \cite{dev:burlington}.}\label{fig:qem}
\end{figure}

We briefly compare results from general error mitigation (GEM) and Qiskit error mitigation (QEM). We limit our comparison to circuits with two qubits. Results obtained using QEM are shown in Fig.~\ref{fig:qem}. They can be directly compared to those shown in Fig.~\ref{fig:2} for which GEM was used. The improvement in error mitigation using GEM, in contrast to QEM, is substantial.

\subsubsection{Negative mitigation}\label{sec1}
Negative mitigation refers to the cases in which the device output is better than the mitigated output. Although Table~\ref{table} shows that a majority of experiments indicate positive mitigation, cases of negative mitigation are also present. In Fig.~\ref{fig:3}, it was observed that experiments with numbers ranging between 1 and 40 correspond to negative mitigation and are comparatively large in number. Owing to this, we restrict our discussion about cases with negative mitigations to the experiments shown in Fig.~\ref{fig:3}, but it also applies to circuits with other qubit numbers and mean depths. 

For a two-qubit circuit there are only four possible output states. Measurement outputs thereof, range from all normalised frequencies in only one of the states to frequencies equally distributed over all states. As it is reasonable to expect more errors as the depth of a circuit increases, the actual rate depends on various factors, inter alia, the device performance. With an increasing number of errors, the device generates data that deviates more and more from the theoretical results, ultimately producing output states having almost the same frequency ($\approx$ 0.25). Thus, if we intended to have an output equally distributed ($= 0.25$) over all states, errors will not drive the frequencies of the output states too far away from the intended ones, as compared to cases that have only one state as output. When we wish to mitigate errors in the former case, and use $C_c$ that give independent states, the device will be unable to perform well. Then, $M_G$ will not be able to mitigate errors significantly because the errors, although present, did not produce sufficient error effects (relevant for GEM) in the first place. By inspecting the simulation results (not shown), we see that almost all experiments that show negative mitigation have one thing in common, namely, that their circuit produces output states with equal probabilities. Note that the generated circuits were random, and no control was exercised over what the outcomes should be.

From the aforementioned discussion and all the available experimental data, it is discernible that negative mitigation mostly appears in cases where the device is giving outputs $V$ close to the expected ones $E$ (i.e.~small $\Delta V$). In such cases mitigation may not be required at all. Since GEM is a post-processing method, this flexibility offers a user discretion over the need of its use. 

Future works need to address the ability of GEM to predict beforehand the cases where mitigation is not required, thereby saving resources. In such undertakings, $M_G$ may be helpful.

\subsubsection{Calibration circuits}\label{sec:rc}
Here we address the question: Is it possible to prepare different calibration circuits? We saw that GEM as a method is able to mitigate errors, given that some assumptions are satisfied. We defined $M_G$ = $(M_1+M_2)/2$ as the final calibration matrix. Before examining other possibilities, let us briefly discuss the motivations to do so.

If we want to reproduce errors occurring in a given circuit $C_g$, we should (in the ideal case) use $C_g$ also as the calibration circuit(s), i.e. $C_c\approx C_g$. We use the approximate sign to indicate that the number of gates in both circuits is almost the same and it originates from the fact that we sometimes need an extra gate for the state preparation in $C_c$, and so the depth may increase. Such an approach, however, is not possible because we do not know a priori the outputs of $C_g$. A circuit similar to $C_g$ for which the outputs are always known is simply $C_{2c}\approx C_g^\dagger C_g$ (identity circuit). This doubles the depth that is why we name it $C_{2c}$. Thus, by using this approach we are forced to approximate the errors produced in $C_g$ of depth $D$ with errors produced in $C_{2c}$ of depth $2D$. Working with  noisy intermediate scale quantum (NISQ) devices like those of IBM, we find that circuits with different depths exhibit different errors. Then, doubling the depth is likely to make the assumption of Sect.~\ref{sec:qemod} inapplicable. For this reason, we cut $C_{g}$ in half and then take $C_c \approx C_{g/2}^\dagger C_{g/2}$ in order to keep the depth(s) of the calibration circuit(s) $C_c$ and the given circuit $C_g$ very close.

\begin{figure}
\begin{tikzpicture}\centering
\node at (-1,-0.9) {\textbf{$\ket{00}$}};
\fill[blue!40] (0,-0.3) rectangle (0.6, -0.9);
\node[draw] at (0.3,-0.6) {$R_x$};
\draw (-0.5,-1.3) -- (0,-1.3);
\draw (2.3,-1.3) -- (1.6,-1.3);
\draw (0.0,-1.3) -- (1.6,-1.3);

\fill[green!60] (2,-1) rectangle (2.6, -1.6);
\draw (2.6,-1.3) -- (2.9,-1.3);
\node at (3.0,-1.3) {\textbf{.}};
\node at (3.3,-1.3) {\textbf{.}};
\node at (3.6,-1.3) {\textbf{.}};

\node[draw] at (2.3,-1.3) {\textbf{Y}};
\fill[red!60] (4.0,-1.0) rectangle (4.6, -1.6);
\node at (4.3,-1.3) {\textbf{$M_z$}};

\draw (0.6,-0.6) -- (1,-0.6);
\draw (-0.5,-0.6) -- (0,-0.6);
\draw (2.3,-0.6) -- (1.6,-0.6);
\fill[green!60] (1,-0.3) rectangle (1.6,- 0.9);

\draw (2.3,-0.6) -- (2.9,-0.6);
\node at (3.0,-0.6) {\textbf{.}};
\node at (3.3,-0.6) {\textbf{.}};
\node at (3.6,-0.6) {\textbf{.}};

\node[draw] at (1.3,-0.6) {\textbf{X}};
\draw (1.8,-0.6) -- (1.8,-1.3);

\fill[red!60] (4.0,-0.3) rectangle (4.6, -0.9);
\node at (4.3,-0.6) {\textbf{$M_z$}};

\draw (1.8,-1.3) circle (2pt);

\end{tikzpicture} \caption{Circuit ($C_g$) showing a \textbf{$R_x$} gate of $\pi /6$ on qubit 1. The gate sequence \{\textbf{X} gate on qubit 1, \textbf{CNOT} gate with control at qubit 1 and target at qubit 2, and \textbf{Y} gate on qubit 2\} is repeated 30 times followed by a measurement in Z basis.\label{fig:(I)}}
\end{figure}
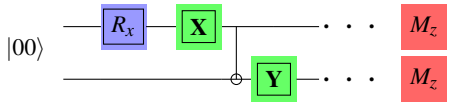

\begin{figure}
\begin{tikzpicture}
\node[inner sep=0pt] (russell) at (0,0)
    {\includegraphics[trim = {0.55cm 0.9cm 0.5cm 0.5cm},scale=0.555]{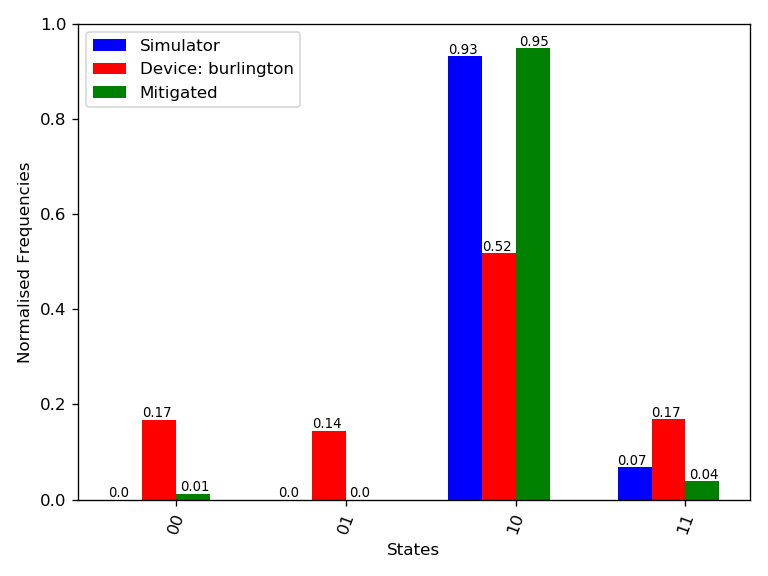}};
\fill [color=white] (-2.1,2.625) rectangle (-1.05,2.27) ;
   \node at (-1.57,2.46) {\scriptsize Burlington};
\end{tikzpicture}
\caption{Simulator (left; blue), device (middle; red), and mitigated (right; green) results for the circuit shown in Fig.~\ref{fig:(I)}. The device used was IBM Q Burlington \cite{dev:burlington}.}\label{fig:method1}
\end{figure}

Other possible calibration circuit(s) seem to be cases where $C_g$ produces (known) output(s) which are linearly independent. In such cases we could directly use $C_c\approx C_g$. While GEM can be applied to any circuit, it is still always possible to use any other $C_c$, the gates in which could be completely different from those in $C_g$. However, we cannot expect $\Delta_G$ for such cases to be positive. We now come back to the question raised above, and answer it in the affirmative, by proposing the following two methods:
\begin{enumerate}[(I)]
 \item If the outputs of $C_g$ are known (or could be known) to be linearly independent, the best strategy may be to use $C_c \approx C_g$.
 \item If a set (or sets) of gates (say $\{g_1,g_2,...,g_p\}$) in $C_g$ are known to produce some errors which can be reproduced by another set of gates (say $\{g^c_1 ,g^c_2 \,...,g_q^c\}$), such that $q<p$, we may reduce the depth of $C_c$ to be less than depth of $C_g$ by an amount $q-p$.
\end{enumerate}
To show the flexibility of GEM, we present two simple examples where (I) or (II) may help. Consider the two-qubit circuit shown in Fig.~\ref{fig:(I)}.
We find that because of the presence of the $R_x$ gate in $C_g$, calibration circuits do not (theoretically) give linearly independent states in the output. Now, since the number of all other gates is large, the error contribution from $R_x$ may be ignored and we can remove it altogether from $C_c$. Doing so will enable us to fill all the columns of $M$ with circuits that produce independent states. Note that by following this new procedure, we abandon requirement 7. The results for this example are given in Fig.~\ref{fig:method1}. We see that the mitigated data are close to the theoretical (simulator) results. The calibration matrix for this experiment: 
\begin{equation}
M = 
\begin{pmatrix}
0.5526123 & 0.1893310 & 0.1623535 & 0.1437988 \\
0.1372070 & 0.5322266 & 0.1494141 & 0.1748047 \\
0.1693115 & 0.1330566 & 0.5349121 & 0.1687012 \\
0.1408691 & 0.1453857 & 0.1533203 & 0.5126953
\end{pmatrix}.\label{eqn:mat}
\end{equation}

In cases where the total number of gates which drive the circuit output in linearly dependent states (like the $R_x$ gate in the previous example) is significant in comparison with the total number of gates, method (II) may help. As an example, we consider the circuit in Fig.~\ref{fig:(I)} but we replace each \textbf{Y} gate by an $R_y$ gate. In such a case, if the errors produced by, say, thirty $R_y$ gates, can be satisfactorily reproduced by, say, ten \textbf{Z} gates, the thirty $R_y$ gates may be replaced by ten \textbf{Z} gates, placed anywhere on qubit 2, under the assumption that an erroneous gate operation produces errors independent of its position in the circuit. 

If we restrict ourselves to a device and a limited `universal' gate-set, and pre-identify gates that closely reproduce the errors produced by multiple other gates, error mitigation may be made quicker. Future work is necessary to address this issue.

\subsubsection{Calibration matrix}
The calibration matrix plays a central role in both QEM and GEM. Both the methods differ insofar the ways they design the circuits that fill the matrix. This matrix can also offer other insights, as follows. QEM and GEM can work for arbitrary long circuits if the device produces distinguishable column entries in the calibration matrix. This can be seen by observing the entries within each column of the matrix. If the entries are all nearly equal, then the device was working completely randomly and the output was uniform over all states, and we cannot expect positive mitigation. Alternatively, if the matrix's columns contain distinguishable entries, error mitigation is possible, as seen from the matrix in Eq.~(\ref{eqn:mat}) and the corresponding positive mitigation in Fig.~\ref{fig:method1}.

\subsubsection{Reducing resource consumption}\label{sec:rrc}
We now look at requirement number 4 as listed in Sect.~\ref{sec:req}, and ask if it is possible to implement GEM using fewer resources. GEM (QEM) requires $2^{N+1}$ ($2^N$) calibration circuits to be run when applied to $N$ qubits, in order to fill the matrix $M_G$ $(M_Q)$. As postulated, this $M_G$ contains all information about the errors that a device is prone to. We may choose to approximate the errors to be reproduced in $C_c$ circuits using circuit set $\{C_1,C_2,...,C_p\}$, where $p<2^{N+1}$ and $c=2^{N+1}$. This will give us a matrix $M_G^*$, which will have $c-p$ unfilled columns, to be filled either using the available information from  $p$ circuits or with ones on the diagonal and zeros on the off-diagonal elements. 

Naturally, such a process may lead to a reduction in cases of significant positive mitigation. We now have fewer empirical entries in $M_G^*$ than in $M_G$, thereby reducing the information we have about the errors in a given circuit. Thus, there is a trade-off between reducing the resource consumption and increasing the mitigation efficiency. As we use fewer resources, mitigation efficiency decreases, and vice versa.

A possible direction for future work could be to find optimal trade-off criteria when resource consumption is extremely high.

\section{Conclusions}
We proposed requirements regarding an ideal error mitigation method. Thereafter, we outlined a general error mitigation method applicable to any quantum circuit. This method was tested on different quantum devices, with one to four qubits, using randomly generated circuits. The circuit depths in these tests were limited by the device performance/fidelity. The results showed significant error mitigation. We discussed possibilities to further improve the method. 

\section{Acknowledgement}
We thank D. Willsch and M. Willsch for detailed discussions and C. D. Gonzalez Calaza for help with Qiskit. We acknowledge use of the IBM Q for this work. The views expressed are those of the authors and do not reflect the official policy or position of IBM or the IBM Q team. M.S.J.  acknowledges support from the project OpenSuperQ (820363) of the EU Quantum Flagship.
\bibliographystyle{apsrev4-1}  
\bibliography{references}  

\end{document}